\documentclass{ws-ijmpa}
\usepackage[super,compress]{cite}
\def\Journal#1#2#3#4{{#1} {\bf #2}, #3 (#4)}

\def\NIMA{\em Nucl. Instrum. Methods A}

\def\PLB{\em Phys. Lett.  B}
\def\PRL{\em Phys. Rev. Lett.}
\def\PRD{\em Phys. Rev. D}

\def\GaC{\em Gravitation and Cosmology}
\def\GaCS{\em Gravitation and Cosmology Suppl.}

\def\PAN{\em Phys.Atom.Nucl.}

\def\APJ{\em Astrophys. J.}
\def\SCI{\em Science}
\def\MPLA{\em Mod. Phys. Lett. A}
\def\IJTP{\em Int. J. Theor. Phys.}
\def\IJMPA{\em Int. J. Mod. Phys. A}
\def\IJMPD{\em Int. J. Mod. Phys.  D}
\def\NJP{\em New J. of Phys.}
\def\ARAA{\em Ann. Rev. Astron. Astrophys.}
\def\AIPCP{\em AIP Conf. Proc.}
\def\JHEP{\em JHEP}
\def\JCAP{\em JCAP}

\def\JPCS{\em J. Phys. Conf. Ser.}
\def\BWP{\em Bled Workshops in Physics}
\def\s{{\,\rm s}}
\def\g{{\,\rm g}}
\def\eV{\,{\rm eV}}
\def\keV{\,{\rm keV}}
\def\MeV{\,{\rm MeV}}
\def\GeV{\,{\rm GeV}}
\def\TeV{\,{\rm TeV}}
\def\sv{\left<\sigma v\right>}
\def\({\left(}
\def\){\right)}
\def\cm{{\,\rm cm}}
\def\K{{\,\rm K}}
\def\kpc{{\,\rm kpc}}
\def\beq{\begin{equation}}
\def\eeq{\end{equation}}
\def\bea{\begin{eqnarray}}
\def\eea{\end{eqnarray}}
\begin{document}

\markboth{M.Yu.Khlopov}
{Dark Atoms and Puzzles of Dark Matter Searches}

%
\catchline{}{}{}{}{}
%

\title{DARK ATOMS AND PUZZLES OF DARK MATTER SEARCHES
}

\author{MAXIM KHLOPOV}

\address{Centre for Cosmoparticle Physics "Cosmion"  \\
National Research Nuclear University "Moscow Engineering Physics Institute", 115409 Moscow, Russia \\
APC laboratory 10, rue Alice Domon et L\'eonie Duquet \\75205
Paris Cedex 13, France\\
khlopov@apc.univ-paris7.fr}

\maketitle

\begin{history}
\received{Day Month Year}
\revised{Day Month Year}
\end{history}

\begin{abstract}
The nonbaryonic dark matter of the Universe is assumed to consist of
new stable forms of matter. Their stability reflects symmetry of micro world and particle candidates for cosmological
dark matter are the lightest particles that bear new conserved quantum numbers. Dark matter candidates can appear in the new families of quarks and leptons and the existence of new stable charged leptons and quarks is possible, if they are hidden in elusive "dark atoms". Such possibility, strongly restricted by the constraints on anomalous isotopes of light elements, is not excluded in scenarios that predict stable double charged particles.
The excessive -2 charged particles are bound in these scenarios with
primordial helium in O-helium "atoms", maintaining specific
nuclear-interacting form of the dark matter, which may
provide an interesting solution for the puzzles of the direct dark matter searches.
\keywords{elementary particles; dark matter; early universe.}
\end{abstract}

\ccode{PACS numbers:12.60.Cn,98.90.+s,12.60.Nz,14.60.Hi,26.35.+c,36.90.+f,03.65.Ge}
\section{Introduction}

 Extensions of the standard model
imply new symmetries and new particle states. In particle theory Noether's theorem relates the exact symmetry to
conservation of respective charge. If the symmetry is strict, the charge is strictly conserved. The lightest particle, bearing this charge, is
stable. It gives rise to the deep relationship between dark matter
candidates and particle symmetry beyond the Standard model.

According to the modern cosmology, the dark matter, corresponding to
$\sim 25\%$ of the total cosmological density, is nonbaryonic and
consists of new stable forms of matter. These forms of matter (see e.g. Refs. \citen{book,newBook,DMRev}
for review and reference) should
be stable, saturate the measured dark matter density and decouple
from plasma and radiation at least before the beginning of matter
dominated stage. The easiest way to satisfy these conditions is to
involve neutral elementary weakly interacting particles. However it
is not the only particle physics solution for the dark matter
problem and more evolved models of the physical nature of dark matter are
possible.

Formation of the Large Scale Structure of the Universe from
small initial density fluctuations is one of the most
important reasons for the {\it nonbaryonic} nature of the dark matter
that is decoupled from matter and radiation and provides the effective growth of these fluctuations before recombination. It implies dark matter candidates from the physics beyond the Standard model (see Refs. \citen{DMRev,LSSFW,Gelmini,Aprile:2009zzd,Feng:2010gw} for recent review). On the other hand, the initial density fluctuations, coming from the very early Universe are also originated from physics beyond the Standard model. In the present review we give some examples, linking the primordial seeds of galaxy formation to effects of particle symmetry breaking at very high energies.

Here we don't touch the exciting problems of the possible nature of dark matter related with extra dimensions and brane cosmology, but even in the case of our 1+3 dimensional space-time we find a lot of examples of nontrivial candidates for cosmological dark matter.

In the Section \ref{pattern} we present examples of various types of particle candidates for dark matter. We then pay special attention to a possibility for stable charged species of new quarks and leptons to form dark matter, hidden in neutral dark atoms (Section \ref{asymmetry}). In Section \ref{ohe} we consider specific form of O-helium (OHe) dark atoms that consist of heavy -2 charged heavy lepton-like particle surrounded by helium nuclear shell. The qualitative advantages of this OHe scenario and the problems of its proof on the basis of a strict quantum mechanical solution of the problem of OHe interaction with nuclei are discussed in Section \ref{solution}. The conclusive Section \ref{Discussion} considers the challenges for experimental test of the OHe solution for the puzzles of dark matter searches.

\section{Particle physics candidates for dark matter}
 \label{pattern}
Most of the known particles are unstable. For a particle with the
mass $m$ the particle physics time scale is $t \sim 1/m$
\footnote{Here and further, if it isn't specified otherwise we use the units $\hbar=c=k=1$}, so in
particle world we refer to particles with lifetime $\tau \gg 1/m$
as to metastable. To be of cosmological significance in the Big Bang Universe metastable
particle should survive after the temperature of the Universe $T$
fell down below $T \sim m$, what means that the particle lifetime
should exceed $t \sim (m_{Pl}/m) \cdot (1/m)$. Such a long
lifetime should find reason in the existence of an (approximate)
symmetry. From this viewpoint, cosmology is sensitive to the most
fundamental properties of microworld, to the conservation laws
reflecting strict or nearly strict symmetries of particle theory.

So, electron is absolutely
stable, what reflects the conservation of electric charge. In the
same manner the stability of proton is conditioned by the
conservation of baryon charge. The stability of ordinary matter is
thus protected by the conservation of electric and baryon charges,
and its properties reflect the fundamental physical scales of
electroweak and strong interactions. Indeed, the mass of electron
is related to the scale of the electroweak symmetry breaking,
whereas the mass of proton reflects the scale of QCD confinement.

The new strict symmetry
is then reflected in the existence of new stable particles, which
should be present in the Universe and considered as candidates for dark matter.

\subsection{Stable relics. Freezing out. Charge symmetric case}
 \label{WIMPs}
The simplest form of dark matter candidates is the gas of new
stable neutral massive particles, originated from early Universe. For
particles with the mass $m$, at high temperature $T>m$ the
equilibrium condition, $$n \cdot \sigma v \cdot t > 1$$ is valid, if
their annihilation cross section $\sigma > 1/(m m_{Pl})$ is
sufficiently large to establish the equilibrium. At $T<m$ such
particles go out of equilibrium and their relative concentration
freezes out. This
is the main idea of calculation of primordial abundance for
Weakly Interacting Massive Particles (WIMPs, see e.g. Refs.
\citen{book,newBook,DMRev} for details).

The process of WIMP annihilation to ordinary particles, considered in $t$channel,
determines their scattering cross section on ordinary particles and thus
relates the primordial abundance of WIMPs to their scattering rate in the
ordinary matter. Forming nonluminous massive halo of our Galaxy, WIMPs can penetrate
the terrestrial matter and scatter on nuclei in underground detectors. The strategy of
direct WIMP searches implies detection of recoil nuclei from this scattering.

The process inverse to annihilation of WIMPs corresponds to their production in collisions
of ordinary particles. It should lead to effects of missing mass and energy-momentum,
being the challenge for experimental search for production of dark matter candidates at accelerators,
e.g. at LHC.

\subsection{Stable relics. Decoupling}

More weakly interacting and/or more light species decouple from plasma
and radiation being relativistic
at $T \gg m$, when $$n \cdot \sigma v \cdot t \sim 1,$$
i.e. at $$T_{dec} \sim (\sigma m_{Pl})^{-1} \gg m.$$ After decoupling these species retain
their equilibrium distribution until they become non-relativistic at $T < m$.
Conservation of partial entropy in the cosmological expansion links the modern abundance
of these species to number density of relic photons with the account for the increase of
the photon number density due to the contribution of heavier ordinary particles, which were
in equilibrium in the period of decoupling.

For the long time, it seemed possible that relic neutrinos can be the dominant form of cosmological dark matter and the corresponding neutrino-dominated Universe was considered as physical ground of Hot Dark Matter scenario of Large scale structure formation.  Experimental discovery of neutrino oscillations together with stringent upper limits on the mass of electron neutrino exclude this possibility. Moreover, even neutrino masses in the range of 1eV lead to features in the spectrum of density fluctuations that are excluded by the observational data of CMB.

\subsection{Stable relics. SuperWIMPs}
The maximal
temperature, which is reached in inflationary Universe, is the
reheating temperature, $T_{r}$, after inflation. So, the very
weakly interacting particles with the annihilation cross section
$$\sigma < 1/(T_{r} m_{Pl}),$$ as well as very heavy particles with
the mass $$m \gg T_{r}$$ can not be in thermal equilibrium, and the
detailed mechanism of their production should be considered to
calculate their primordial abundance.

In particular, thermal production of gravitino in very early Universe is proportional to the reheating temperature $T_{r}$, what puts upper limit on this temperature from constraints on primordial gravitino abundance\cite{khlopovlinde,khlopovlinde2,khlopovlinde3,khlopov3,khlopov31,Karsten,Kawasaki}.
\subsection{Axions and axion-like particles}
A wide class of particle models possesses a symmetry breaking
pattern, which can be effectively described by
pseudo-Nambu--Goldstone (PNG) field (see Refs. \citen{DMRev,book2,PBHrev} for review and references). The coherent oscillations of this field represent a specific type
of CDM in spite of a very small mass of PNG particles $m_a=\Lambda^2/f$, where $f \gg \Lambda$, since these particles are created in Bose-Einstein condensate in the ground state, i.e. they are initially created as nonrelativistic in the very early Universe.
This feature, typical for invisible axion models can be the general feature for all the axion-like PNG particles.

At high temperatures the pattern of successive spontaneous and manifest breaking of global U(1) symmetry implies the
succession of second order phase transitions. In the first
transition at $T \sim f$, continuous degeneracy of vacua leads, at scales
exceeding the correlation length, to the formation of topological
defects in the form of a string network; in the second phase
transition at $T \sim \Lambda \ll f$, continuous transitions in space between degenerated
vacua form surfaces: domain walls surrounded by strings. This last
structure is unstable, but, as was shown in the example of the
invisible axion \cite{Sakharov2,kss,kss2}, it is reflected in the
large scale inhomogeneity of distribution of energy density of
coherent PNG (axion) field oscillations. This energy density is
proportional to the initial value of phase, which acquires dynamical
meaning of amplitude of axion field, when axion mass $m_a=C m_{\pi}f_{\pi}/f$ (where $m_{\pi}$ and $f_{\pi}\approx m_{\pi}$ are the pion mass and constant, respectively, the constant $C\sim 1$ depends on the choice of the axion model and $f\gg f_{\pi}$ is the scale of the Peccei-Quinn symmetry breaking) is switched on
in the result of the second phase transition.

The value of phase changes by $2 \pi$ around string. This strong
nonhomogeneity of phase leads to corresponding nonhomogeneity of
energy density of coherent PNG (axion) field oscillations. Usual
argument (see e.g. Ref. \citen{kim} and references therein) is essential
only on scales, corresponding to mean distance between strings.
This distance is small, being of the order of the scale of
cosmological horizon in the period, when PNG field oscillations
start. However, since the nonhomogeneity of phase follows the
pattern of axion string network this argument misses large scale
correlations in the distribution of oscillations' energy density.

Indeed, numerical analysis of string network (see review in the
Ref. \citen{vs}) indicates that large string loops are strongly suppressed
and the fraction of about 80\% of string length, corresponding to
long loops, remains virtually the same in all large scales. This
property is the other side of the well known scale invariant
character of string network. Therefore the correlations of energy
density should persist on large scales, as it was revealed in Refs.
\citen{Sakharov2,kss,kss2}. Discussion of such primordial inhomogeneous structures of dark matter
go beyond the scope of the present paper and we can recommend the interested reader
Refs. \citen{DMRev,book2,PBHrev} for review and references.
\subsection{Self interacting dark matter}\label{mirror}
Extensive hidden sector of particle theory can provide the existence of new interactions, which only new particles possess. Historically one of the first examples of such self-interacting dark matter was presented by the model of mirror matter. Mirror particles, first proposed by T. D. Lee and C. N. Yang in Ref. \citen{LeeYang} to restore equivalence of left- and right-handed co-ordinate systems in the presence of P- and C- violation in weak interactions, should be strictly symmetric by their properties to their ordinary twins. After discovery of CP-violation it was shown by I. Yu. Kobzarev, L. B. Okun and I. Ya. Pomeranchuk in Ref. \citen{KOP} that mirror partners cannot be associated with antiparticles and should represent a new set of symmetric partners for ordinary quarks and leptons with their own strong, electromagnetic and weak mirror interactions. It means that there should exist mirror quarks, bound in mirror nucleons by mirror QCD forces and mirror atoms, in which mirror nuclei are bound with mirror electrons by mirror electromagnetic interaction \cite{ZKrev,FootVolkas}. If gravity is the only common interaction for ordinary and mirror particles, mirror matter can be present in the Universe in the form of elusive mirror objects, having symmetric properties with ordinary astronomical objects (gas, plasma, stars, planets...), but causing only gravitational effects on the ordinary matter \cite{Blin1,Blin2}.

Even in the absence of any other common interaction except for gravity, the observational data on primordial helium abundance and upper limits on the local dark matter seem to exclude mirror matter, evolving in the Universe in a fully symmetric way in parallel with the ordinary baryonic matter\cite{Carlson,FootVolkasBBN}. The symmetry in cosmological evolution of mirror matter can be broken either by initial conditions\cite{zurabCV,zurab}, or by breaking mirror symmetry in the sets of particles and their interactions as it takes place in the shadow world\cite{shadow,shadow2}, arising in the heterotic string model. We refer to Refs.
\citen{newBook,OkunRev,Paolo} for current review of mirror matter and its cosmology.

If new particles possess new $y$-charge, interacting with massless bosons or intermediate bosons with sufficiently small mass ($y$-interaction),  for slow $y$-charged particles Coulomb-like factor
of "Gamov-Sommerfeld-Sakharov enhancement" \cite{Som,Sak,Sakhenhance} should be added in
the annihilation cross section
$$C_y=\frac{2 \pi \alpha_y/v}{1 - \exp{(-2 \pi \alpha_y/v)}},$$
where $v$ is relative velocity and $\alpha_y$ is the running gauge constant of $y$-interaction. This factor may not be essential in the period of particle freezing out in the early Universe (when $v$ was only few times smaller than $c$, but can cause strong enhancement in the effect of annihilation of nonrelativistic dark matter particles in the Galaxy.
\subsection{Subdominant dark matter}
If charge symmetric stable particles (and their antiparticles) represent
only subdominant fraction of the cosmological dark matter, more detailed analysis
of their distribution in space, of their condensation in galaxies,
of their capture by stars, Sun and Earth, as well as effects of
their interaction with matter and of their annihilation provides
more sensitive probes for their existence.

In particular,
hypothetical stable neutrinos of 4th generation with mass about 50
GeV should be the subdominant form of modern dark
matter, contributing less than 0,1 \% to the total density
\cite{ZKKC,DKKM}. However, direct experimental search for cosmic
fluxes of weakly interacting massive particles (WIMPs) may be
sensitive to existence of such component (see Refs.
\citen{DAMA,DAMA-review,Bernabei:2008yi,DAMA2,DAMA3,DAMA4,CDMS,CDMS2,CDMS3,CDMSi,xenon,lux,cogent} and references therein). It was
shown in Refs. \citen{Fargion99,Grossi,Belotsky,Belotsky2} that
annihilation of 4th neutrinos and their antineutrinos in the Galaxy
is severely constrained by the measurements of gamma-background, cosmic positrons
and antiprotons. 4th neutrino
annihilation inside the Earth should lead to the flux of underground
monochromatic neutrinos of known types, which can be traced in the
analysis of the already existing and future data of underground
neutrino detectors \cite{Belotsky,BKS1,BKS2,BKS3}.

An interesting multi-component scenario, based on millicharges and presented in Ref.
\citen{Wallemacq:2013hsa}, proposes a dark sector composed of traditional collisionless
particles and of a subdominant more complex part, where two new kinds
of fermions are introduced and form hydrogen-like atoms through a
dark $U\left(1\right)$ gauge coupling carried out by a dark massless
photon. While one of the two species is light and plays the role of
a dark electron, the other one is heavy and is seen as the nucleus
of the atom. The latter is coupled to a dark neutral scalar via a
Yukawa coupling, creating a finite-range attraction between dark nuclei.
Non-gravitational interactions between the dark and the ordinary sectors
come into play through the kinetic and mass mixings between the photon
and the dark photon and between the standard $\sigma$ meson and the
dark scalar respectively. These have straightforward consequences
in direct-dark-matter-search experiments since both dark fermions
have small effective couplings to the standard photon while the dark
nucleus is coupled to the $\sigma$ meson, making it capable of interacting
with nucleons. The dark atoms of the halo, that might form a disk,
hit the surface of the Earth and collide with terrestrial atoms until
they lose all their energy and thermalize. This happens before they
reach an underground detector, typically located at a depth of $1$
km, after what they start sinking down, driven by gravity, and arrive
in the detector with thermal energies. This makes it impossible to
produce nuclear recoils but the dark nuclei bind to the nuclei of
the active medium via radiative capture, which causes the emission
of photons that produce the observed signal. The model, thanks to
its complex subdominant part, can reproduce well the results from
DAMA/LIBRA and CoGeNT without contradicting with the negative results
from XENON100, LUX and CDMS-II/Ge.
\subsection{Decaying dark matter}
Decaying particles with lifetime $\tau$, exceeding the age of the
Universe, $t_{U}$, $\tau > t_{U}$, can be treated as stable. By
definition, primordial stable particles survive to the present time
and should be present in the modern Universe. The net effect of
their existence is given by their contribution into the total
cosmological density. However, even small effect of their decay
can lead to significant contribution to cosmic rays and gamma background\cite{ddm}.
Leptonic decays of dark matter are considered as possible explanation of
the cosmic positron excess, measured in the range above 10 GeV by PAMELA\cite{pamela}, FERMI/LAT\cite{lat} and AMS02\cite{ams2}.

Primordial unstable particles with the lifetime, less than the age
of the Universe, $\tau < t_{U}$, can not survive to the present
time. But, if their lifetime is sufficiently large to satisfy the
condition $\tau \gg (m_{Pl}/m) \cdot (1/m)$, their existence in
early Universe can lead to direct or indirect traces\cite{khlopov7}.

Weakly interacting particles, decaying to invisible modes, can influence Large Scale Structure formation.
Such decays prevent formation of the structure, if they take place before the structure is formed.
Invisible products of decays after the structure is formed should contribute in the cosmological dark energy.
The Unstable Dark matter scenarios\cite{Sakharov1,UDM,UDM1,UDM2,UDM3,berezhiani4,berezhiani5,TSK,GSV} implied weakly interacting particles that form the structure on the matter dominated stage and then decay to invisible modes after the structure is formed.

Cosmological
flux of decay products contributing into the cosmic and gamma ray
backgrounds represents the direct trace of unstable particles\cite{khlopov7,sedelnikov}. If
the decay products do not survive to the present time their
interaction with matter and radiation can cause indirect trace in
the light element abundance\cite{khlopovlinde3,khlopov3,khlopov31,DES} or in the fluctuations of thermal
radiation\cite{UDM4}.
\subsection{Charge asymmetry of dark matter}
The fact that particles are not absolutely stable means that the corresponding charge is not strictly conserved and generation particle charge asymmetry is possible, as it is assumed for ordinary baryonic matter. At sufficiently strong particle annihilation cross section excessive particles (antiparticles) can dominate in the relic density, leaving exponentially small admixture of their antiparticles (particles) in the same way as primordial excessive baryons dominate over antibaryons in baryon asymmetric Universe. In this case {\it Asymmetric dark matter} doesn't lead to significant effect of particle annihilation in the modern Universe and can be searched for either directly in underground detectors or indirectly by effects of decay or condensation and structural transformations of e.g. neutron stars (see Ref. \citen{adm} for recent review and references). If particle annihilation isn't strong enough, primordial pairs of particles and antiparticles dominate over excessive particles (or antiparticles) and this case has no principle difference from the charge symmetric case. In particular, for very heavy charged leptons (with the mass above 1 TeV), like "tera electrons"\cite{Glashow}, discussed in \ref{asymmetry}, their annihilation due to electromagnetic interaction is too weak to provide effective suppression of primordial tera electron-positron pairs relative to primordial asymmetric excess\cite{BKSR1}.
\subsection{Charged stable relics. Dark atoms}
New particles with electric charge and/or strong interaction can
form anomalous atoms and contain in the ordinary matter as anomalous
isotopes. For example, if the lightest quark of 4th generation is
stable, it can form stable charged hadrons, serving as nuclei of
anomalous atoms of e.g. anomalous helium
\cite{BKSR1,BKS,BKSR,FKS,I,BKSR4}. Therefore, stringent upper limits on anomalous isotopes, especially, on anomalous hydrogen put severe constraints on the existence of new stable charged particles. However, as we discuss in the rest of this review, stable doubly charged particles can not only exist, but even dominate in the cosmological dark matter, being effectively hidden in neutral "dark atoms"\cite{DADM}.

\section{Stable charged constituents of Dark Atoms}\label{asymmetry}
New stable particles may possess new U(1)
gauge charges and bind by Coulomb-like forces in composite dark
matter species. Such dark atoms cannot be luminous, since they
radiate invisible light of U(1) photons. Historically mirror matter
(see subsubsection \ref{mirror} and Refs. \citen{book,OkunRev} for review and references) seems to be the
first example of such an atomic dark matter.

However, it turned out that the possibility of new stable electrically charged leptons and quarks is not completely excluded and Glashow's tera-helium\cite{Glashow} has offered a new solution for this type of
dark atoms of dark matter. Tera-U-quarks with electric charge +2/3
formed stable (UUU) +2 charged "clusters" that formed with two -1
charged tera-electrons E neutral [(UUU)EE] tera-helium "atoms" that
behaved like Weakly Interacting Massive Particles (WIMPs). The main
problem for this solution was to suppress the abundance of
positively charged species bound with ordinary electrons, which
behave as anomalous isotopes of hydrogen or helium. This problem
turned to be unresolvable\cite{BKSR1}, since the model\cite{Glashow}
predicted stable tera-electrons $E^-$ with charge -1.
As soon as primordial helium is formed in the Standard Big Bang
Nucleosynthesis (SBBN) it captures all the free $E^-$ in positively
charged $(He E)^+$ ion, preventing any further suppression of
positively charged species. Therefore, in order to avoid anomalous
isotopes overproduction, stable particles with charge -1 (and
corresponding antiparticles) should be absent, so that stable
negatively charged particles should have charge -2 only.

Elementary particle frames for heavy stable -2 charged species are
provided by: (a) stable "antibaryons" $\bar U \bar U \bar U$ formed
by anti-$U$ quark of fourth generation\cite{Q,I,BKSR4,Belotsky:2008se,DADM}
(b) AC-leptons\cite{DADM,FKS}, predicted in the
extension \cite{FKS} of standard model, based on the approach of
almost-commutative geometry\cite{bookAC}.  (c) Technileptons and
anti-technibaryons \cite{KK} in the framework of walking technicolor
models (WTC)\cite{Sannino:2004qp,Hong:2004td,Dietrich:2005jn,Dietrich:2005wk,Gudnason:2006ug,Gudnason:2006yj}. (d) Finally, stable charged
clusters $\bar u_5 \bar u_5 \bar u_5$ of (anti)quarks $\bar u_5$ of
5th family can follow from the approach, unifying spins and charges\cite{Norma,Norma2,Norma3,Norma4,Norma5}. Since all these models also predict corresponding +2
charge antiparticles, cosmological scenario should provide mechanism
of their suppression, what can naturally take place in the
asymmetric case, corresponding to excess of -2 charge species,
$O^{--}$. Then their positively charged antiparticles can
effectively annihilate in the early Universe.

If new stable species belong to non-trivial representations of
electroweak SU(2) group, sphaleron transitions at high temperatures
can provide the relationship between baryon asymmetry and excess of
-2 charge stable species, as it was demonstrated in the case of WTC
in Refs. \citen{KK,Levels1,KK2,unesco,iwara,I2}.


\subsection{Problem of tera-fermion composite dark matter}
Glashow's Tera-helium Universe was first inspiring example of the
composite dark matter scenario. $SU(3)_c \times SU(2) \times SU(2)'
\times U(1)$ gauge model\cite{Glashow} was aimed to explain the origin of the neutrino mass and to solve the problem of strong CP-violation in QCD. New extra $SU(2)'$ symmetry acts on three heavy generations of
tera-fermions  linked with the light fermions by $CP'$
transformation. $SU(2)'$ symmetry breaking at TeV scale makes
tera-fermions much heavier than their light partners. Tera-fermion
mass spectrum is the same as for light generations, but all the
masses are scaled by the same factor of about $10^6$. Thus the
masses of lightest heavy particles are in {\it tera}-eV (TeV) range,
explaining their name.

Glashow's model\cite{Glashow} takes into account
that
 very heavy quarks $Q$ (or antiquarks $\bar Q$) can form bound states with other heavy quarks
 (or antiquarks) due to their Coulomb-like QCD attraction, and the binding energy of these states
 substantially exceeds the binding energy of QCD confinement.
Then stable $(QQq)$ and $(QQQ)$ baryons can exist.

According to Ref. \citen{Glashow} primordial heavy quark $U$ and heavy
electron $E$ are stable and
may form a neutral $(UUUEE)$ "atom"
with $(UUU)$ hadron as nucleus and two $E^-$s as "electrons". The
gas of such "tera-helium atoms" was proposed in Ref. \citen{Glashow} as a candidate for a
WIMP-like dark matter.

The problem of such scenario is an
inevitable presence of "products of incomplete combustion" and the
necessity to decrease their abundance.

Unfortunately, as it was shown in Ref. \citen{BKSR1}, this
picture of Tera-helium Universe can not be realized.

When ordinary $^4$He is formed in Big Bang
Nucleosynthesis, it binds all the free
$E^-$ into positively charged $(^4HeE^-)^+$ "ions". This puts
Coulomb barrier for any successive $E^-E^+$ annihilation or any
effective $EU$ binding. It removes  a possibility to suppress the abundance of
unwanted tera-particle species (like $(eE^+)$, $(^4He Ee)$ etc).
For instance the remaining abundance of $(eE^+)$ and $(^4HeE^-e)$ exceeds the terrestrial upper limit for anomalous hydrogen by
{\it 27 orders} of magnitude\cite{BKSR1}.

\subsection{Composite dark matter from almost commutative geometry}
The AC-model is based on the specific mathematical approach of
unifying general relativity, quantum mechanics and gauge symmetry\cite{FKS,bookAC}.
This realization naturally embeds the Standard model, both
reproducing its gauge symmetry and Higgs mechanism with prediction of a Higgs boson mass. AC model
 is in some sense alternative to SUSY, GUT and superstring extension of Standard model. The AC-model\cite{FKS} extends the fermion content of the Standard
model by two heavy particles, $SU(2)$ electro-weak singlets, with opposite electromagnetic charges.
Each of them has its own antiparticle. Having no other gauge charges of Standard model,
these particles (AC-fermions) behave as heavy stable leptons with
charges $-2e$ and $+2e$, called $A^{--}$ and $C^{++}$, respectively.

Similar to the Tera-helium Universe, AC-lepton relics from
intermediate stages of a multi-step process towards a final $(AC)$
atom formation must survive in the present Universe. In spite of the assumed excess of
particles ($A^{--}$ and $C^{++}$) the abundance of relic
antiparticles ($\bar A^{++}$ and $\bar C^{--}$) is not negligible.
There may be also a significant fraction of $A^{--}$ and $C^{++}$, which remains
unbound after recombination process of these particles into $(AC)$ atoms took place. As soon as $^4He$ is formed in Big
Bang nucleosynthesis, the primordial component of free anion-like AC-leptons
($A^{--}$) is mostly trapped in the first three minutes into a
neutral O-helium atom $^4He^{++}A^{--}$.
O-helium is able to capture free $C^{++}$ creating $(AC)$ atoms and releasing $^4He$ back. In the same way the annihilation of antiparticles speeds up. $C^{++}$-O-helium reactions stop, when their timescale exceeds a cosmological time, leaving O-helium and $C^{++}$ relics in the Universe. The catalytic reaction of O-helium with $C^{++}$ in the dense matter bodies provides successive
$(AC)$ binding that suppresses terrestrial
anomalous isotope abundance below the experimental upper limit. Due to screened charge of AC-atoms they have WIMP-like interaction with the ordinary matter. Such WIMPs are inevitably accompanied by a tiny component of nuclear interacting O-helium.

\subsection{Stable charged techniparticles in Walking Technicolor}

The minimal walking technicolor model\cite{Sannino:2004qp,Hong:2004td,Dietrich:2005jn,Dietrich:2005wk,Gudnason:2006ug,Gudnason:2006yj}
has two techniquarks, i.e. up $U$ and down $D$, that transform
under the adjoint representation of an $SU(2)$ technicolor gauge
group. The six
Goldstone bosons $UU$, $UD$, $DD$ and their corresponding
antiparticles carry technibaryon number since they are made of
two techniquarks or two anti-techniquarks. This means that if there is no
processes violating the technibaryon number the lightest
technibaryon will be stable.

The electric charges of $UU$, $UD$,
and $DD$ are given in general by $q+1$, $q$, and $q-1$
respectively, where $q$ is an arbitrary real number. The model requires in addition
the existence of a fourth family of leptons, i.e. a ``new
neutrino'' $\nu'$ and a ``new electron'' $\zeta$. Their electric charges are in
terms of $q$ respectively $(1-3q)/2$ and $(-1-3q)/2$.

There are three possibilities for a scenario of dark atoms of dark matter. The first one is to have an excess of $\bar{U}\bar{U}$ (charge $-2$).
The technibaryon
number $TB$ is conserved and therefore $UU$ (or $\bar{U}\bar{U}$) is
stable. The second possibility is to
have excess of $\zeta$ that also has $-2$ charge and is
stable, if $\zeta$ is lighter than $\nu'$ and technilepton number $L'$  is conserved. In the both cases
stable particles with $-2$ electric charge have substantial relic
densities and can capture $^4He^{++}$ nuclei to form a neutral techni-O-helium
atom.
Finally there is a
possibility to have both  $L'$ and $TB$ conserved. In this case, the dark matter would be composed
of bound atoms $(^4He^{++}\zeta^{--})$ and $(\zeta^{--}(U U )^{++})$. In the latter case the excess of $\zeta^{--}$ should be larger, than the excess of $(U U )^{++})$, so that WIMP-like $(\zeta^{--}(U U )^{++})$ is subdominant at the dominance of nuclear interacting techni-O-helium.

The technicolor and the
Standard Model particles are in thermal equilibrium as long as the
timescale of the weak (and color) interactions is smaller than the
cosmological time. The sphalerons allow violation of  $TB$, of baryon number $B$, of lepton number $L$ and  $L'$ as
long as the temperature of the Universe exceeds the electroweak scale.
It was shown in\cite{KK} that there is a balance between the excess of techni(anti)baryons, $(\bar{U}\bar{U})^{--}$,
technileptons $\zeta^{--}$ or of the both over the corresponding particles ($UU$ and/or $\zeta^{++}$) and the observed baryon asymmetry
of the Universe. It was also shown the there are parameters of the model, at which this asymmetry has
proper sign and value, explaining the dark matter density.

\subsection{\label{4generation} Stable particles of 4th generation matter}
Modern precision data
on the parameters of the Standard model do not exclude\cite{Maltoni:1999ta}
the existence of
the  4th generation of quarks and leptons. The 4th generation follows from heterotic string phenomenology and
its difference from the three known light generations can be
explained by a new conserved charge, possessed only by
its quarks and leptons\cite{Q,I,Belotsky:2000ra,Belotsky:2005uj,Belotsky:2004st}. Strict conservation of this charge makes the
lightest particle of 4th family (neutrino) absolutely
stable, but it was shown in Refs. \citen{Belotsky:2000ra,Belotsky:2005uj,Belotsky:2004st} that this neutrino cannot be the dominant form of the dark matter.
The same conservation law requires the lightest quark to be long living
\cite{Q,I}. In principle the lifetime of $U$ can exceed the age of the
Universe, if $m_U<m_D$\cite{Q,I}.
Provided that sphaleron transitions establish excess of $\bar U$ antiquarks at the observed baryon asymmetry
 $(\bar U \bar U \bar U)$ can be formed and bound with $^4He$ in atom-like state
of O-helium\cite{I}.

In the successive discussion of OHe dark matter we generally don't specify the type of $-2$ charged particle, denoting it as $O^{--}$.
However, one should note that the AC model doesn't provide OHe as the dominant form of dark matter, so that the quantitative features of OHe dominated Universe are not related to this case.
\section{Dark atoms with helium shell}
\label{ohe}
Here we concentrate on the properties of OHe atoms, their interaction with matter and qualitative picture of OHe cosmological evolution\cite{I,Levels,FKS,KK,unesco,Khlopov:2008rp,KhlopovPHE} and observable effects. We show following Refs. \citen{DADM,DMDA} that interaction of OHe with nuclei in
underground detectors can  explain positive results
of dark matter searches in DAMA/NaI (see for review Ref. \citen{DAMA-review})
and DAMA/LIBRA\cite{Bernabei:2008yi}
experiments by annual modulations of radiative capture of O-helium, resolving the controversy
between these results and the results of other experimental groups.

After it is formed
in the Standard Big Bang Nucleosynthesis (SBBN), $^4He$ screens the excessive
$O^{--}$ charged particles in composite $(^4He^{++}O^{--})$ {\it
O-helium} ($OHe$) ``atoms''\cite{I}.

In all the considered forms of O-helium, $O^{--}$ behaves either as lepton or
as specific "heavy quark cluster" with strongly suppressed hadronic
interaction. Therefore O-helium interaction with matter is
determined by nuclear interaction of $He$. These neutral primordial
nuclear interacting species can play the role of a nontrivial form of strongly
interacting dark matter\cite{Starkman,Wolfram,Starkman2,Javorsek,Mitra,Mack,McGuire:2001qj,McGuire2,ZF}, giving rise to a Warmer than
Cold dark matter scenario\cite{Levels,Levels1,KK2}.
\subsection{OHe atoms and their interaction with nuclei}
The structure of OHe atom follows from the general
analysis of the bound states of $O^{--}$ with nuclei.

Consider a simple model\cite{Cahn,Pospelov,Kohri}, in which the nucleus is
regarded as a sphere with uniform charge density and in which the
mass of the $O^{--}$ is assumed to be much larger than that of the
nucleus. Spin dependence is also not taken into account so that both
the particle and nucleus are considered as scalars. Then the
Hamiltonian is given by
\begin{equation}
    H=\frac{p^2}{2 A m_p} - \frac{Z Z_x \alpha}{2 R} + \frac{Z Z_x \alpha}{2 R} \cdot (\frac{r}{R})^2,
\end{equation}
for short distances $r<R$ and
\begin{equation}
    H=\frac{p^2}{2 A m_p} - \frac{Z Z_x \alpha}{R},
\end{equation}
for long distances $r>R$, where $\alpha$ is the fine structure
constant, $R = d_o A^{1/3} \sim 1.2 A^{1/3} /(200 MeV)$ is the
nuclear radius, $Z$ is the electric charge of nucleus and $Z_x=2$ is
the electric charge of negatively charged particle $X^{--}$. Since
$A m_p \ll M_X$ the reduced mass is $1/m= 1/(A m_p) + 1/M_X \approx
1/(A m_p)$.

For small nuclei the Coulomb binding energy is like in hydrogen atom
and is given by
\begin{equation}
    E_b=\frac{1}{2} Z^2 Z_x^2 \alpha^2 A m_p.
\end{equation}

For large nuclei $X^{--}$ is inside nuclear radius and the harmonic
oscillator approximation is valid for the estimation of the binding
energy
\begin{equation}
    E_b=\frac{3}{2}(\frac{Z Z_x \alpha}{R}-\frac{1}{R}(\frac{Z Z_x \alpha}{A m_p R})^{1/2}).
\label{potosc}
\end{equation}

For the intermediate regions between these two cases with the use of
trial function of the form $\psi \sim e^{- \gamma r/R}$ variational
treatment of the problem\cite{Cahn,Pospelov,Kohri} gives
\begin{equation}
    E_b=\frac{1}{A m_p R^2} F(Z Z_x \alpha A m_p R ),
\end{equation}
where the function $F(a)$ has limits
\begin{equation}
    F(a \rightarrow 0) \rightarrow \frac{1}{2}a^2  - \frac{2}{5} a^4
\end{equation}
and
\begin{equation}
    F(a \rightarrow \infty) \rightarrow \frac{3}{2}a  - (3a)^{1/2},
\end{equation}
where $a = Z Z_x \alpha A m_p R$. For $0 < a < 1$ the Coulomb model
gives a good approximation, while at $2 < a < \infty$ the harmonic
oscillator approximation is appropriate.

In the case of OHe $a = Z Z_x \alpha A m_p R \le 1$, what proves its
Bohr-atom-like structure, assumed in Refs. \citen{I,KK,unesco,iwara,I2}.
The radius of Bohr orbit in these ``atoms"
\cite{I,Levels} $r_{o} \sim 1/(Z_{o} Z_{He}\alpha m_{He}) \approx 2
\cdot 10^{-13} \cm $.
However, the size of
He nucleus, rotating around $O^{--}$ in this Bohr atom, turns out to be of
the order and even a bit larger than the radius $r_o$ of its Bohr
orbit, and the corresponding correction to the binding energy due to
non-point-like charge distribution in He is significant.

Bohr atom like structure of OHe seems to provide a possibility to
use the results of atomic physics for description of OHe interaction
with matter. However, the situation is much more complicated. OHe
atom is similar to the hydrogen, in which electron is hundreds times
heavier, than proton, so that it is proton shell that surrounds
"electron nucleus". Nuclei that interact with such "hydrogen" would
interact first with strongly interacting "protonic" shell and such
interaction can hardly be treated in the framework of perturbation
theory. Moreover in the description of OHe interaction the account
for the finite size of He, which is even larger than the radius of
Bohr orbit, is important. One should consider, therefore, the
analysis, presented below, as only a first step approaching true
nuclear physics of OHe.

The approach of Refs. \citen{Levels,Levels1} assumes the following
picture of OHe interaction with nuclei: OHe is a neutral atom in the ground state,
perturbed  by Coulomb and nuclear forces of the approaching nucleus.
The sign of OHe polarization changes with the distance: at larger distances Stark-like effect takes place - nuclear Coulomb force polarizes OHe so that  nucleus is attracted by the induced dipole moment of OHe, while as soon as the perturbation by nuclear force starts to dominate the nucleus polarizes OHe in the opposite way so that He is situated more close to the nucleus, resulting in the repulsive effect of the helium shell of OHe.
When helium is completely merged with the nucleus the interaction is
reduced to the oscillatory potential of $O^{--}$ with
homogeneously charged merged nucleus with the charge $Z+2$.

\begin{figure}
    \begin{center}
        \includegraphics[scale=0.3]{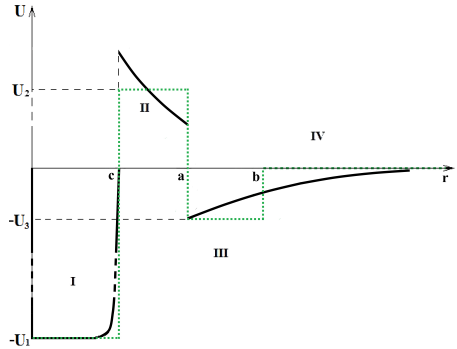}
        \caption{The potential of OHe-nucleus system and its rectangular well approximation.}
        \label{pic1}
    \end{center}
\end{figure}

\begin{figure}
\begin{center}
\includegraphics[scale=0.6]{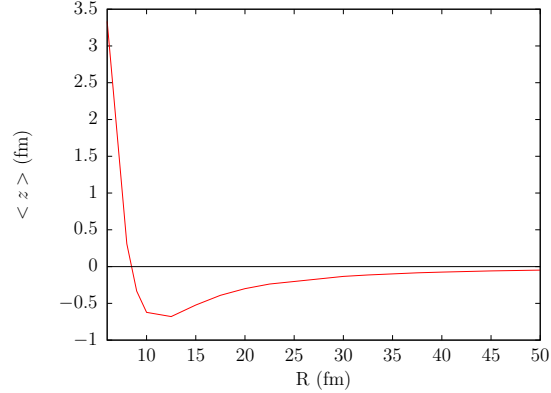}
\caption{Polarization $<z>$ (Fm) of OHe as a function of the distance
$R$ (fm) of an external sodium nucleus, calculated in Ref. \citen{quentin} in the framework of perturbation theory.}
\label{Pol}
 \end{center}
\end{figure}

Therefore OHe-nucleus potential can have qualitative feature, presented on Fig.~\ref{pic1}:
the potential well at large distances (regions III-IV) is changed by a potential wall in region II. The existence of this potential barrier is crucial for all the qualitative features of OHe scenario: it causes suppression of reactions with transition of OHe-nucleus system to levels in the potential well of the region I, provides the dominance of elastic scattering while transitions to levels in the shallow well (regions III-IV) should dominate in reactions of OHe-nucleus capture. The proof of this picture implies accurate and detailed quantum-mechanical treatment, which was started in Ref. \citen{quentin}. With the use of perturbation theory it was shown that OHe polarization changes sign, as the nucleus approaches OHe (as it is given on Fig. \ref{Pol}), but the perturbation approach was not valid for the description at smaller distances, while the estimations indicated that this change of polarization may not be sufficient for creation of the potential, given by Fig.~\ref{pic1}. If the picture of Fig.~\ref{pic1} is not proved, one may need more sophisticated models retaining the ideas of OHe scenario, which involve more elements of new physics, as proposed in Ref. \citen{Wallemacq:2013hsa}.

On the other hand, O-helium, being an $\alpha$-particle with screened electric charge,
can catalyze nuclear transformations, which can influence primordial
light element abundance and cause primordial heavy element
formation. It is especially important for quantitative estimation of
role of OHe in Big Bang Nucleosynthesis and in stellar evolution.
These effects need a special detailed and complicated
study and
this work is under way.

The qualitative
picture of OHe cosmological evolution is presented below following Refs. \citen{DMRev,I,Levels,FKS,KK,Levels1,unesco,Khlopov:2008rp,DADM}
and is based on the idea of the dominant role of elastic
collisions in OHe interaction with baryonic matter.

\subsection{Large Scale structure formation by OHe dark matter}
Due to elastic nuclear interactions of its helium constituent with nuclei in
the cosmic plasma, the O-helium gas is in thermal equilibrium with
plasma and radiation on the Radiation Dominance (RD) stage, while
the energy and momentum transfer from plasma is effective. The
radiation pressure acting on the plasma is then transferred to
density fluctuations of the O-helium gas and transforms them in
acoustic waves at scales up to the size of the horizon.

At temperature $T < T_{od} \approx 1 S_3^{2/3}\eV$ the energy and
momentum transfer from baryons to O-helium is not effective
\cite{I,KK} because $$n_B \sv (m_p/m_o) t < 1,$$ where $m_o$ is the
mass of the $OHe$ atom and $S_3= m_o/(1 \TeV)$. Here \beq \sigma
\approx \sigma_{o} \sim \pi r_{o}^2 \approx
10^{-25}\cm^2\label{sigOHe}, \eeq and $v = \sqrt{2T/m_p}$ is the
baryon thermal velocity. Then O-helium gas decouples from plasma. It
starts to dominate in the Universe after $t \sim 10^{12}\s$  at $T
\le T_{RM} \approx 1 \eV$ and O-helium ``atoms" play the main
dynamical role in the development of gravitational instability,
triggering the large scale structure formation. The composite nature
of O-helium determines the specifics of the corresponding dark
matter scenario.

At $T > T_{RM}$ the total mass of the $OHe$ gas with density $\rho_d
= (T_{RM}/T) \rho_{tot} $ is equal to
$$M=\frac{4 \pi}{3} \rho_d t^3 = \frac{4 \pi}{3} \frac{T_{RM}}{T} m_{Pl}
(\frac{m_{Pl}}{T})^2$$ within the cosmological horizon $l_h=t$. In
the period of decoupling $T = T_{od}$, this mass  depends strongly
on the O-helium mass $S_3$ and is given by \cite{KK}\beq M_{od} =
\frac{T_{RM}}{T_{od}} m_{Pl} (\frac{m_{Pl}}{T_{od}})^2 \approx 2
\cdot 10^{44} S^{-2}_3 \g = 10^{11} S^{-2}_3 M_{\odot}, \label{MEPm}
\eeq where $M_{\odot}$ is the solar mass. O-helium is formed only at
$T_{o}$ and its total mass within the cosmological horizon in the
period of its creation is $M_{o}=M_{od}(T_{od}/T_{o})^3 = 10^{37}
\g$.

On the RD stage before decoupling, the Jeans length $\lambda_J$ of
the $OHe$ gas was restricted from below by the propagation of sound
waves in plasma with a relativistic equation of state
$p=\epsilon/3$, being of the order of the cosmological horizon and
equal to $\lambda_J = l_h/\sqrt{3} = t/\sqrt{3}.$ After decoupling
at $T = T_{od}$, it falls down to $\lambda_J \sim v_o t,$ where $v_o
= \sqrt{2T_{od}/m_o}.$ Though after decoupling the Jeans mass in the
$OHe$ gas correspondingly falls down
$$M_J \sim v_o^3 M_{od}\sim 3 \cdot 10^{-14}M_{od},$$ one should
expect a strong suppression of fluctuations on scales $M<M_o$, as
well as adiabatic damping of sound waves in the RD plasma for scales
$M_o<M<M_{od}$. It can provide some suppression of small scale
structure in the considered model for all reasonable masses of
O-helium. The significance of this suppression and its effect on the
structure formation needs a special study in detailed numerical
simulations. In any case, it can not be as strong as the free
streaming suppression in ordinary Warm Dark Matter (WDM) scenarios,
but one can expect that qualitatively we deal with Warmer Than Cold
Dark Matter model.

At temperature $T < T_{od} \approx 1 S_3^{2/3} \keV$ the energy and
momentum transfer from baryons to O-helium is not effective\cite{I,Levels,Levels1}
and O-helium gas decouples from plasma. It
starts to dominate in the Universe after $t \sim 10^{12}\s$  at $T
\le T_{RM} \approx 1 \eV$ and O-helium ``atoms" play the main
dynamical role in the development of gravitational instability,
triggering the large scale structure formation. The composite nature
of O-helium determines the specifics of the corresponding warmer than cold dark
matter scenario.

Being decoupled from baryonic matter, the $OHe$ gas does not follow
the formation of baryonic astrophysical objects (stars, planets,
molecular clouds...) and forms dark matter halos of galaxies. It can
be easily seen that O-helium gas is collisionless for its number
density, saturating galactic dark matter. Taking the average density
of baryonic matter one can also find that the Galaxy as a whole is
transparent for O-helium in spite of its nuclear interaction. Only
individual baryonic objects like stars and planets are opaque for
it.

\subsection{Anomalous component of cosmic rays}
O-helium atoms can be destroyed in astrophysical processes, giving
rise to acceleration of free $O^{--}$ in the Galaxy.

O-helium can be ionized due to nuclear interaction with cosmic rays\cite{I,I2}.
Estimations\cite{I,Mayorov} show that for the number
density of cosmic rays $ n_{CR}=10^{-9}\cm^{-3}$ during the age of
Galaxy a fraction of about $10^{-6}$ of total amount of OHe is
disrupted irreversibly, since the inverse effect of recombination of
free $O^{--}$ is negligible. Near the Solar system it leads to
concentration of free $O^{--}$ $ n_{O}= 3 \cdot 10^{-10}S_3^{-1}
\cm^{-3}.$ After OHe destruction free $O^{--}$ have momentum of
order $p_{O} \cong \sqrt{2 \cdot m_{o} \cdot I_{o}} \cong 2 \GeV
S_3^{1/2}$ and velocity $v/c \cong 2 \cdot 10^{-3} S_3^{-1/2}$ and
due to effect of Solar modulation these particles initially can
hardly reach Earth\cite{KK2,Mayorov}. Their acceleration by Fermi
mechanism or by the collective acceleration forms power spectrum of
$O^{--}$ component at the level of $O/p \sim n_{O}/n_g = 3 \cdot
10^{-10}S_3^{-1},$ where $n_g \sim 1 \cm^{-3}$ is the density of
baryonic matter gas.

At the stage of red supergiant stars have the size $\sim 10^{15}
\cm$ and during the period of this stage$\sim 3 \cdot 10^{15} \s$,
up to $\sim 10^{-9}S_3^{-1}$ of O-helium atoms per nucleon can be
captured\cite{KK2,Mayorov}. In the Supernova explosion these OHe
atoms are disrupted in collisions with particles in the front of
shock wave and acceleration of free $O^{--}$ by regular mechanism
gives the corresponding fraction in cosmic rays. However, this
picture needs detailed analysis, based on the development of OHe
nuclear physics and numerical studies of OHe evolution in the
stellar matter.

If these mechanisms of $O^{--}$ acceleration are effective, the
anomalous low $Z/A$ component of $-2$ charged $O^{--}$ can be
present in cosmic rays at the level $O/p \sim n_{O}/n_g \sim
10^{-9}S_3^{-1},$ and be within the reach for PAMELA and AMS02
cosmic ray experiments.

In the framework of Walking Technicolor model the excess of both
stable $\zeta^{--}$ and $(UU)^{++}$ is possible\cite{KK2}, the latter
being two-three orders of magnitude smaller, than the former. It
leads to the two-component composite dark matter scenario with the
dominant OHe accompanied by a subdominant WIMP-like component of
$(\zeta^{--}(U U )^{++})$ bound systems. Technibaryons can
be metastable and decays of $(UU)^{++}$ can provide
explanation for anomalies, observed in high energy cosmic positron
spectrum by PAMELA, FERMI-LAT and AMS02.

\subsection{Positron annihilation and gamma lines in galactic
bulge}
Inelastic interaction of O-helium with the matter in the
interstellar space and its de-excitation can give rise to radiation
in the range from few keV to few  MeV. In the galactic bulge with
radius $r_b \sim 1 \kpc$ the number density of O-helium can reach
the value $n_o\approx 3 \cdot 10^{-3}/S_3 \cm^{-3}$ and the
collision rate of O-helium in this central region was estimated in Refs.
\citen{I2,KK2}: $dN/dt=n_o^2 \sigma v_h 4 \pi r_b^3 /3 \approx 3 \cdot
10^{42}S_3^{-2} \s^{-1}$. At the velocity of $v_h \sim 3 \cdot 10^7
\cm/\s$ energy transfer in such collisions is $\Delta E \sim 1 \MeV
S_3$. These collisions can lead to excitation of O-helium. If nS ($n \ge3$)
level is excited, pair production dominates over two-photon channel
in the de-excitation by $E0$ transition and positron production with
the rate $3 \cdot 10^{42}S_3^{-2} \s^{-1}$ is not accompanied by
strong gamma signal. According to Ref. \citen{Finkbeiner:2007kk} this rate
of positron production for $S_3 \sim 1$ is sufficient to explain the
excess in positron annihilation line from bulge, measured by
INTEGRAL (see Ref. \citen{integral} for review and references).
The dependence of this effect on the mass of O-helium, as well as on
its density profile and velocity dispersion in the galactic bulge is
studied in Ref. \citen{DApositrons}.

If $OHe$
levels with nonzero orbital momentum are excited, gamma lines should
be observed from transitions ($ n>m$) $E_{nm}= 1.598 \MeV (1/m^2
-1/n^2)$ (or from the similar transitions corresponding to the case
$I_o = 1.287 \MeV $) at the level $3 \cdot 10^{-4}S_3^{-2}(\cm^2 \s
\MeV ster)^{-1}$.
\section{O-helium solution for dark matter puzzles}
\label{solution}
It should be noted that the nuclear cross section of the O-helium
interaction with matter escapes the severe constraints\cite{McGuire:2001qj,McGuire2,ZF}
on strongly interacting dark matter particles
(SIMPs)\cite{Starkman,Wolfram,Starkman2,Javorsek,Mitra,Mack,McGuire:2001qj,McGuire2,ZF} imposed by the XQC experiment\cite{XQC,XQC1}. Therefore, a special strategy of direct O-helium  search
is needed, as it was proposed in Ref. \citen{Belotsky:2006fa}.

\subsection{O-helium in the terrestrial matter} The evident
consequence of the O-helium dark matter is its inevitable presence
in the terrestrial matter, which appears opaque to O-helium and
stores all its in-falling flux.

After they fall down terrestrial surface, the in-falling $OHe$
particles are effectively slowed down due to elastic collisions with
matter. Then they drift, sinking down towards the center of the
Earth with velocity \beq V = \frac{g}{n \sigma v} \approx 80 S_3
A_{med}^{1/2} \cm/\s. \label{dif}\eeq Here $A_{med} \sim 30$ is the average
atomic weight in terrestrial surface matter, $n=2.4 \cdot 10^{24}/A$
is the number of terrestrial atomic nuclei, $\sigma v$ is the rate
of nuclear collisions and $g=980~ \cm/\s^2$.

Near the Earth's surface, the O-helium abundance is determined by
the equilibrium between the in-falling and down-drifting fluxes.

At a depth $L$ below the Earth's surface, the drift timescale is
$t_{dr} \sim L/V$, where $V \sim 400 S_3 \cm/\s$ is the drift velocity and $m_o=S_3 \TeV$ is the mass of O-helium. It means that the change of the incoming flux,
caused by the motion of the Earth along its orbit, should lead at
the depth $L \sim 10^5 \cm$ to the corresponding change in the
equilibrium underground concentration of $OHe$ on the timescale
$t_{dr} \approx 2.5 \cdot 10^2 S_3^{-1}\s$.

The equilibrium concentration, which is established in the matter of
underground detectors at this timescale, is given by
\begin{equation}
    n_{oE}=n_{oE}^{(1)}+n_{oE}^{(2)}\cdot sin(\omega (t-t_0))
    \label{noE}
\end{equation}
with $\omega = 2\pi/T$, $T=1yr$ and
$t_0$ the phase.
So, there is a averaged concentration given by
\begin{equation}
    n_{oE}^{(1)}=\frac{n_o}{320S_3 A_{med}^{1/2}} V_{h}
\end{equation}
and the annual modulation of concentration characterized by the amplitude
\begin{equation}
    n_{oE}^{(2)}= \frac{n_o}{640S_3 A_{med}^{1/2}} V_E.
\end{equation}
Here $V_{h}$-speed of Solar System (220 km/s), $V_{E}$-speed of
Earth (29.5 km/s) and $n_{0}=3 \cdot 10^{-4} S_3^{-1} \cm^{-3}$ is the
local density of O-helium dark matter.

\subsection{OHe in the underground detectors}

The explanation\cite{Levels,DMDA,DADM} of the results of
DAMA/NaI\cite{DAMA-review} and DAMA/LIBRA\cite{Bernabei:2008yi}
experiments is based on the idea that OHe,
slowed down in the matter of detector, can form a few keV bound
state with nucleus, in which OHe is situated \textbf{beyond} the
nucleus. Therefore the positive result of these experiments is
explained by annual modulation in reaction of radiative capture of OHe
\begin{equation}
A+(^4He^{++}O^{--}) \rightarrow [A(^4He^{++}O^{--})]+\gamma
\label{HeEAZ}
\end{equation}
by nuclei in DAMA detector.

To simplify the solution of Schrodinger equation the
potential was approximated in Refs. \citen{Levels,Levels1} by a rectangular potential, presented on Fig.~\ref{pic1}.
Solution of Schrodinger equation determines the condition, under
which a low-energy  OHe-nucleus bound state appears in the shallow well of the region
III and the range of nuclear parameters was found, at which OHe-sodium binding energy is in the interval 2-4 keV.


The rate of radiative capture of OHe by nuclei can be calculated\cite{Levels,DMDA}
with the use of the analogy with the radiative
capture of neutron by proton with the account for: i) absence of M1
transition that follows from conservation of orbital momentum and
ii) suppression of E1 transition in the case of OHe. Since OHe is
isoscalar, isovector E1 transition can take place in OHe-nucleus
system only due to effect of isospin nonconservation, which can be
measured by the factor $f = (m_n-m_p)/m_N \approx 1.4 \cdot
10^{-3}$, corresponding to the difference of mass of neutron,$m_n$,
and proton,$m_p$, relative to the mass of nucleon, $m_N$. In the
result the rate of OHe radiative capture by nucleus with atomic
number $A$ and charge $Z$ to the energy level $E$ in the medium with
temperature $T$ is given by
\begin{equation}
    \sigma v=\frac{f \pi \alpha}{m_p^2} \frac{3}{\sqrt{2}} (\frac{Z}{A})^2 \frac{T}{\sqrt{Am_pE}}.
    \label{radcap}
\end{equation}

Formation of OHe-nucleus bound system leads to energy release of its
binding energy, detected as ionization signal.  In the context of
our approach the existence of annual modulations of this signal in
the range 2-6 keV and absence of such effect at energies above 6 keV
means that binding energy $E_{Na}$ of Na-OHe system in DAMA experiment should
not exceed 6 keV, being in the range 2-4 keV. The amplitude of
annual modulation of ionization signal can reproduce the result of DAMA/NaI and DAMA/LIBRA
experiments for $E_{Na} = 3 \keV$. The
account for energy resolution in DAMA experiments\cite{DAMAlibra}
can explain the observed energy distribution of the signal from
monochromatic photon (with $E_{Na} = 3 \keV$) emitted in OHe
radiative capture.

At the corresponding nuclear parameters there is no binding
of OHe with iodine and thallium\cite{Levels}.

It should be noted that the results of DAMA experiment exhibit also
absence of annual modulations at the energy of MeV-tens MeV. Energy
release in this range should take place, if OHe-nucleus system comes
to the deep level inside the nucleus. This transition implies
tunneling through dipole Coulomb barrier and is suppressed below the
experimental limits.

For the chosen range of nuclear parameters, reproducing the results
of DAMA/NaI and DAMA/LIBRA, the results of Ref. \citen{Levels} indicate that
there are no levels in the OHe-nucleus systems for heavy nuclei. In
particular, there are no such levels in Xe, what
seem to prevent direct comparison with DAMA results in
XENON100 experiment\cite{xenon} or LUX experiment\cite{lux}. The existence of such level in Ge and the comparison with the results of
CDMS\cite{CDMS,CDMS2,CDMS3} and CoGeNT\cite{cogent} experiments need special study. According to Ref. \citen{Levels} OHe should bind with O and Ca, what is of interest for interpretation of the signal, observed in CRESST-II experiment\cite{cresst}.

In the thermal equilibrium OHe capture rate is proportional to the temperature. Therefore it looks
like it is suppressed in cryogenic detectors by a factor of order
$10^{-4}$. However, for the size of cryogenic devices  less, than
few tens meters, OHe gas in them has the thermal velocity of the
surrounding matter and this velocity dominates in the relative velocity of OHe-nucleus system.
It gives the suppression relative to room temperature
only $\sim m_A/m_o$. Then the rate of OHe radiative capture in
cryogenic detectors is given by Eq.(\ref{radcap}), in which room
temperature $T$ is multiplied by factor $m_A/m_o$. Note that in the case of $T=70\K$ in CoGeNT experiment
relative velocity is determined by the thermal velocity of germanium nuclei, what leads to enhancement relative to cryogenic germanium detectors.
\section{Conclusions}
\label{Discussion}
The existence of heavy stable particles is one of the popular solutions for the dark matter problem.
Usually they are considered to be electrically neutral. But potentially dark matter can be formed by
stable heavy charged particles bound in neutral atom-like states by Coulomb attraction.
Analysis of the cosmological data and atomic composition of the Universe gives the constrains
on the particle charge showing that  only $-2$
charged constituents, being trapped by primordial helium
in neutral O-helium states, can avoid the problem of overproduction of the anomalous isotopes of chemical elements, which are severely constrained by observations. Cosmological model of O-helium dark matter
can even explain puzzles of direct dark matter searches.

The proposed explanation is based on the mechanism of low energy
binding of OHe with nuclei. Within the uncertainty of nuclear
physics parameters there exists a range at which OHe binding energy
with sodium is in the interval 2-4 keV. Annual modulation in radiative capture of OHe to
this bound state leads to the corresponding energy release observed
as an ionization signal in DAMA/NaI and
DAMA/LIBRA experiments.


With the account for high sensitivity of the numerical results to
the values of nuclear parameters and for the approximations, made in
the calculations, the presented results can be considered only as an
illustration of the possibility to explain puzzles of dark matter
search in the framework of composite dark matter scenario. An
interesting feature of this explanation is a conclusion that the
ionization signal may
be absent in detectors containing light (e.g. $^3He$) or heavy (e.g. Xe) elements.
Therefore test of results of DAMA/NaI and
DAMA/LIBRA experiments by other experimental groups can become a
very nontrivial task. Recent indications to positive result in the matter of CRESST detector\cite{cresst},
in which OHe binding is expected together with absence of signal in xenon detectors\cite{xenon,lux}, may qualitatively favor the presented approach. For the same chemical content
an order of magnitude suppression in cryogenic detectors can explain why indications to positive effect in
CoGeNT experiment\cite{cogent} can be compatible with the constraints of CDMS/Ge experiment\cite{CDMS3}. The model predicts a possibility of OHe binding with silicon, but this effect should be suppressed at low temperature in CDMS/Si experiment\cite{CDMSi}.

The present explanation contains distinct
features, by which it can be distinguished from
other recent approaches to this problem\cite{Edward,Foot,Feng1,Feng2,Drob,Feldstein:2009tr,Bai,Feldstein:2009np,Fitzpatrick:2010em,Andreas:2010dz,Alves:2010dd,Barger:2010yn,Savage:2010tg,Hooper:2010uy,Chang:2010pr,Chang:2010en,Barger:2010gv,Feldstein:2010su}

An inevitable consequence of the proposed explanation is appearance
in the matter of underground detectors anomalous
superheavy isotopes, having the mass roughly by $m_o$
larger, than ordinary isotopes of the corresponding elements.

It is interesting to note that in the framework of the presented approach
positive result of experimental search for WIMPs by effect of their
nuclear recoil would be a signature for a multicomponent nature of
dark matter. Such OHe+WIMPs multicomponent dark matter scenarios
naturally follow from AC model \cite{FKS} and can be realized in
models of Walking technicolor \cite{KK2}.

Stable $-2$ charge states ($O^{--}$) can be elementary like AC-leptons or technileptons,
or look like technibaryons. The latter, composed of techniquarks, reveal their structure at much higher energy scale and should be produced at LHC as
elementary species. The signature  for AC leptons and techniparticles is unique and distinctive what  allows
to separate them  from other hypothetical exotic particles.

Since simultaneous production of three $U \bar U$ pairs and
their conversion in two doubly charged quark clusters $UUU$
is suppressed, the only possibility to test the
models of composite dark matter from 4th generation in the collider experiments is a search for production of stable hadrons containing single $U$ or $\bar U$ like $Uud$ and $\bar U u$/$\bar U d$.

The presented approach sheds new light on the physical nature of
dark matter. Specific properties of dark atoms and their
constituents are challenging for the experimental search. The
development of quantitative description of OHe interaction with
matter confronted with the experimental data will provide the
complete test of the composite dark matter model. It challenges search for stable double charged particles at accelerators and cosmic rays as direct experimental probe for charged constituents of dark atoms of dark matter.

\section*{Acknowledgments}
I express my gratitude to K.M. Belotsky, J.R. Cudell, D. Fargion, C. Kouvaris, A.G. Mayorov, E. Yu. Soldatov, C. Stephan and Q. Wallemacq for collaboration in obtaining the original results and to J.R. Cudell and Q. Wallemacq for reading the manuscript and discussions.


\end{document}